\documentclass[a4paper]{PoS}
\usepackage{amsmath}

\renewcommand{\a}{\alpha}
\newcommand{\aem}{\hat\alpha_{\rm em}}
\newcommand{\Lag}{\mathcal{L}}
\newcommand{\Lsix}{\Lag^{(6)}}
\newcommand{\Q}{Q}
\newcommand{\hyp}{\mathsf{y}}
\newcommand{\de}{\partial}
\renewcommand{\dag}{\dagger}

\newcommand{\CHqs}{C_{Hq}^{(1)}}
\newcommand{\CHqt}{C_{Hq}^{(3)}}
\newcommand{\CHls}{C_{Hl}^{(1)}}
\newcommand{\CHlt}{C_{Hl}^{(3)}}

\newcommand{\twotwo}{$\bar\psi\psi\to\bar\psi\psi$ }
\newcommand{\twofour}{$\bar\psi\psi\to\bar\psi\psi\bar\psi\psi$ }

\title{Scheming in the SMEFT}

\ShortTitle{Scheming in the SMEFT}

\author{\speaker{Ilaria Brivio}\thanks{Supported by the Villum Fonden and the Danish National Research Foundation (DNRF91).}\\
        Niels Bohr International Academy and Discovery Center, Niels Bohr Institute,
University of Copenhagen, Blegdamsvej 17, DK-2100 Copenhagen, Denmark\\
        E-mail: \email{ilaria.brivio@nbi.ku.dk}}

\abstract{We discuss the constraints on the Standard Model Effective Field Theory inferred from global fits to electroweak data. Special attention is paid to two unconstrained combinations of Wilson coefficients that are present when the analysis is restricted to measurements of \twotwo scatterings. We illustrate how these unconstrained directions arise due to a reparameterization invariance that characterizes \twotwo processes but is not respected in \twofour scatterings. This invariance is independent of the operator basis adopted and of the choice of the input parameters. This is verified comparing the results obtained in the \{$\aem, \hat m_Z, \hat G_F$\} input scheme with those of a \{$\hat m_W, \hat m_Z, \hat G_F$\} scheme.}

\FullConference{The European Physical Society Conference on High Energy Physics\\
		5-12 July, 2017\\
		Venice}

\begin{document}

\section{Introduction}
The Standard Model Effective Field Theory (SMEFT) is a convenient tool for investigating the presence of new physics sectors at an energy scale $\Lambda \gg v$, being $v\simeq246$~GeV the vacuum expectation value of the Standard Model (SM) Higgs field: $\langle H^\dagger H\rangle = v^2/2$. This corresponds to a scenario in which the new exotic states are too massive to be directly produced or studied at the LHC, but indirect evidence for their presence and properties can emerge in the form of anomalies in measured cross-sections or kinematic distributions. The SMEFT provides a systematic and model-independent parameterization of these effects, based only on electroweak (EW) scale assumptions: the SMEFT Lagrangian is structured as a series of $SU(3)_c\times SU(2)_L\times U(1)_Y$ invariant operators constructed out of SM fields\footnote{In particular, the Higgs field is assumed to be a $SU(2)_L$ doublet. This feature distinguishes the SMEFT from the more general Higgs EFT (HEFT, or electroweak chiral Lagrangian).}
 and ordered by their canonical dimension $d$:
\begin{equation}
 \Lag_{\rm SMEFT} = \Lag_{\rm SM} + \Lag^{(5)} + \Lag^{(6)}+\Lag^{(7)} + \dots
 \qquad\text{ with }\quad
 \Lag^{(d)} = \sum_{i} \frac{C_i^{(d)}}{\Lambda^{d-4}} \Q_i^{(d)}\,.
\end{equation} 
Each Lagrangian term $\Lag^{(d)}$ is written as a sum of $d$-dimensional operators $\Q^{(d)}_i$ that form a complete, non-redundant basis. Assuming conservation of the baryon and lepton numbers, the leading beyond-SM effects are described by the Wilson coefficients $C_i$ of the dimension-6 operators. There are {59 + h.c.} independent structures in $\Lsix$ which, assuming an approximate $U(3)^5$ flavor symmetry among fermion generations, corresponds to a total of 69 independent parameters~\cite{Brivio:2017btx}. 
Here we adopt the so-called ``Warsaw'' basis~\cite{Grzadkowski:2010es} for $\Lsix$ and we use a compact notation in which the factor $\Lambda^{-2}$ is implicitly absorbed into the definition of the coefficients.

\section{Global fits to electroweak observables in the SMEFT}
Experimental constraints on $\Lsix$ can be inferred from a standard global fit procedure to selected measurements. A minimal constraining set of observables is constituted, for instance, by near-$Z$-pole LEPI data~\cite{Z-pole}. EFT analyses of this dataset considering the Warsaw basis were carried out, for instance, in Refs.~\cite{Falkowski:2014tna,Berthier:2015oma,Berthier:2015gja}.
Expanding on the method developed in~\cite{Grinstein:1991cd,Han:2004az}, Ref.~\cite{Berthier:2015gja}
considered 103~independent measurements from PEP, PETRA, TRISTAN, SpS, Tevatron, SLAC, LEPI, LEPII and low energy precision data, and combined them into a global fit to the 19 relevant Wilson coefficients:
\begin{equation}\label{relevant_coefficients}
\begin{aligned}
\big\{&
 C_{He}, C_{Hu}, C_{Hd}, \CHls, \CHlt, \CHqs, \CHqt, C_{HWB}, C_{HD},C_{ll}, C_{ee}, C_{le}, \\
 &C_{eu}, C_{ed}, C_{lu}, C_{ld}, C_{lq}^{(1)}, C_{lq}^{(3)}, C_{qe}
 \big\}.
 \end{aligned}
\end{equation}
Although the observables included largely outnumbered the free parameters, two unconstrained directions were found in the fit space, identified as the two null 
eigenvectors of the Fisher matrix {$I_{ij} =\frac12 \frac{\de^2 \chi^2}{\de C_i \de C_j}$}. This result is in agreement with the previous observations in Refs.~\cite{Han:2004az,Grojean:2006nn} and 
it is further confirmed by the results presented in this talk~\cite{Brivio:2017bnu}: the analysis of Ref.~\cite{Berthier:2015gja} is reproduced independently, retaining only the subset of the 31 most constraining observables and spanning a reduced space of 12 Wilson coefficients (first line of Eq.~\eqref{relevant_coefficients}). In this simplified case and neglecting a possible SMEFT theoretical error~\cite{Passarino:2012cb,Berthier:2015oma,Berthier:2015gja,deFlorian:2016spz}, the unconstrained directions are
\begin{subequations}\label{flatalpha}
\begin{align}
w^{\a}_1&= \frac{C_{Hd}}{3}-2C_{HD}+ C_{He}+\frac{\CHls}{2}-\frac{\CHqs}{6}-\frac{2C_{Hu}}{3}-1.29 (\CHqt+ \CHlt)+ 1.64\, C_{HWB},\\
w^{\a}_2&= \frac{C_{Hd}}{3}-2C_{HD}+ C_{He}+\frac{\CHls}{2}-\frac{\CHqs}{6} -\frac{2 C_{Hu}}{3} + 2.16 (\CHqt+ \CHlt)- 0.16\, C_{HWB}.\end{align}
\end{subequations}
Here the superscript $\a$ indicates that the result has been obtained using the set $\{\aem, \hat m_Z, \hat G_F\}$ as input parameters for the EW sector. The presence of unconstrained directions is also tested adopting the alternative scheme $\{\hat m_W, \hat m_Z, \hat G_F\}$, finding that this choice affects only the numerical coefficients of $(\CHqt+\CHlt)$ and $C_{HWB}$ in Eqs.~\eqref{flatalpha}, without altering the conclusions. The expressions for $w_{1,2}^{m_W}$ can be found in Ref.~\cite{Brivio:2017bnu}, together with a detailed discussion of the motivations and implementation of a $\{\hat m_W, \hat m_Z, \hat G_F\}$ input scheme in the SMEFT .

\begin{figure}[t]\centering
\vspace*{-6mm}
 \includegraphics[width=.55\textwidth,trim={0cm 12cm 0cm 0cm},clip]{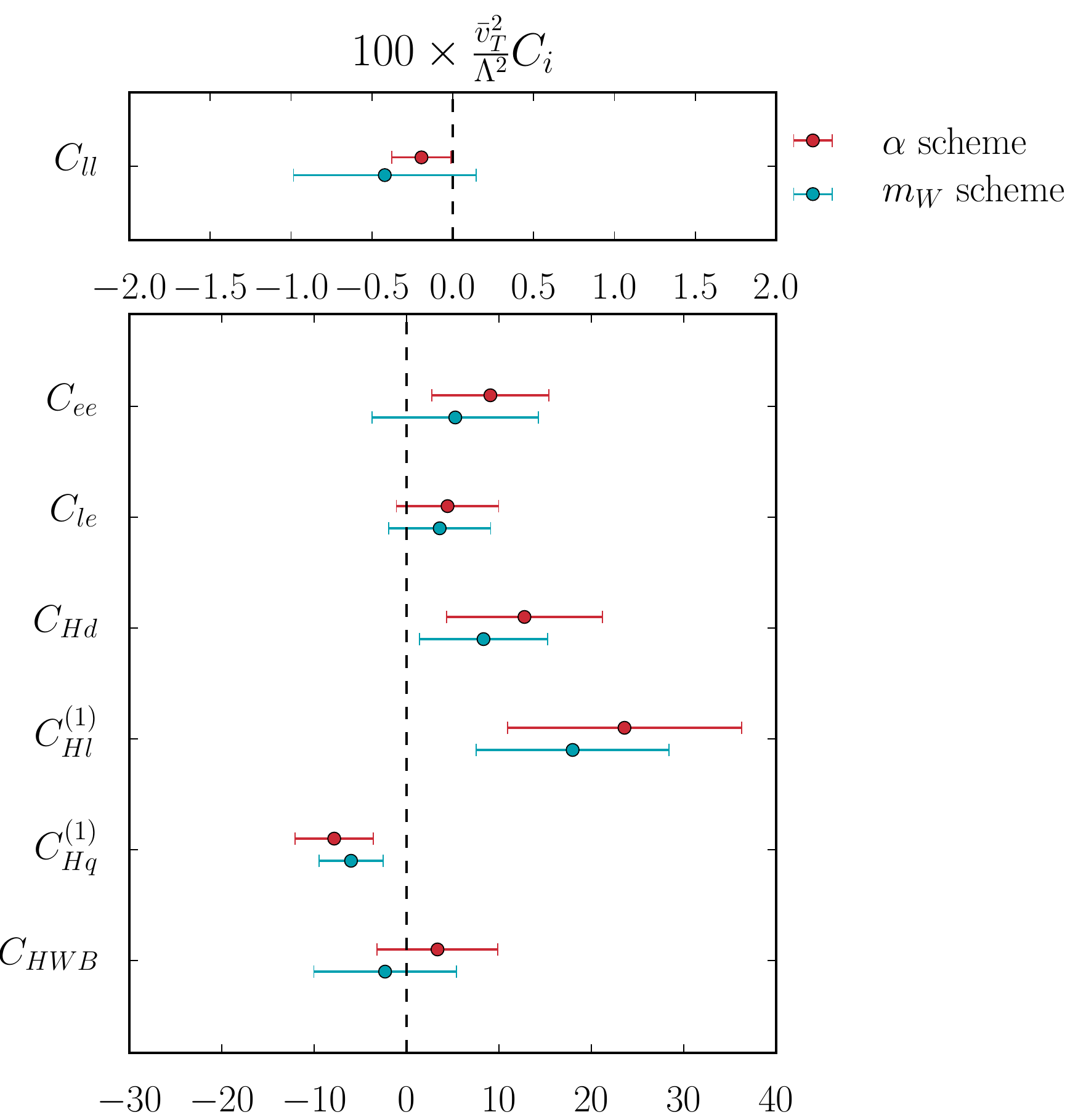}
 \hspace*{-2.cm}
 \includegraphics[width=.4\textwidth,trim={0cm 0cm 4cm 18.5cm},clip]{limits_0err_split2.pdf}
 \caption{Best fit values of the Wilson coefficients (scaled by a factor 100) and corresponding $\pm1\sigma$ confidence regions obtained after profiling away the other parameters~\cite{Brivio:2017bnu}. The fit includes $W^\pm$ pair production data from LEPII and assumes vanishing SMEFT theoretical error.}\label{plot:bounds}
\end{figure}

In order to minimize the prior dependence of the results, confidence intervals for each Wilson coefficient can be extracted from the $\chi^2$ analysis with a profiling method~\cite{Berthier:2015gja}. 
Because the latter requires the Fisher matrix to be invertible, the unconstrained directions have to be lifted to complete the global analysis: a known way to achieve this is by incorporating \twofour scattering data from LEPII to the fit~\cite{Hagiwara:1993ck,DeRujula:1991ufe}.
Following the procedure adopted in Ref.~\cite{Berthier:2016tkq}, the initial dataset of 31 EW observables is thus enlarged with 74 LEPII measurements of \twofour scattering via $W^\pm$ currents~\cite{Brivio:2017bnu}. The confidence regions obtained are shown in Figure~\ref{plot:bounds} for both input schemes. Note that the parameter set has been augmented with $C_W$, that contributes to the anomalous triple gauge couplings $\lambda_{Z,\gamma}$. Remarkably, the fit space is found to be highly correlated, with a mild dependence on the inputs choice, and the strongest correlations appear among the coefficients that participate in the directions $w_{1,2}^\a$ and $C_W$.

\section{Unconstrained directions and reparameterization invariance}
\begin{figure}[t]\centering
\vspace*{-5mm}
\includegraphics[width=.3\textwidth]{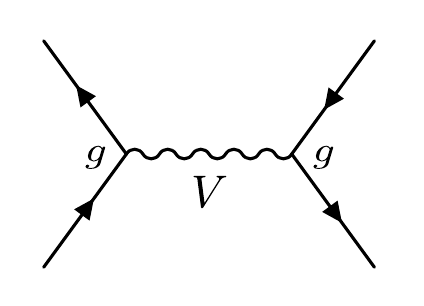}\hspace*{2cm}
\includegraphics[width=.3\textwidth]{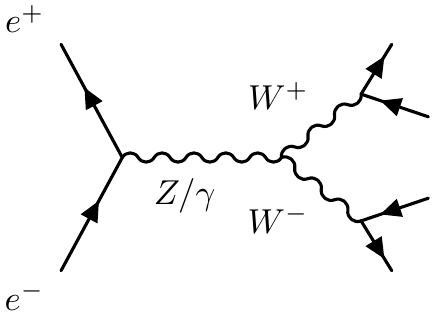}
\caption{Left: Feynman diagram for \twotwo processes. Right: example of diagram contributing to doubly resonant \twofour scattering with charged currents. }\label{fig:diagrams}
\end{figure}

The results described in the previous section point to a peculiar structure of the fit space that, as shown in~\cite{Brivio:2017bnu}, can be explained in terms of a reparameterization invariance present in \twotwo processes, such as those used to extract LEPI data, but not respected in \twofour scatterings. 

Consider a \twotwo process, mediated by a generic vector boson $V$ with an associated coupling constant $g$ (see Fig.~\ref{fig:diagrams}, left).
The relevant Lagrangian reads
\begin{equation}
\label{schematicvector}
\mathcal{L}_{V \psi_i} =  - \frac{1}{4}V^{\mu \, \nu} V_{\mu \, \nu}+ \frac{1}{2} \, m_V^2 \, V^\mu \, V_\mu - g \,\kappa_{ij}\, \bar{\psi}_i \gamma^\mu \psi_j V_\mu + \cdots,
\end{equation}
where $V^{\mu \, \nu} = \partial^\mu \, V^\nu - \partial^\nu \, V^\mu$ and $i,j$ are flavor indices. At tree level and in the limit of massless external fermions, the scattering $\bar\psi\psi\to V \to \bar\psi\psi$ is invariant under the transformation
\begin{equation}\label{Vg_transf}
\left\{\begin{aligned}
V_\mu &\to V'_\mu\,(1+\epsilon) \\
g &\to g'\,/(1+\epsilon)\simeq g'\,(1-\epsilon)
\end{aligned}
\right.
\qquad\quad \text{with }\quad\epsilon \sim \mathcal{O}(v^2/\Lambda^2),
\end{equation} 
that brings the kinetic term of the $V$ field to a non-canonical form without altering the $V\bar\psi\psi$ couplings. This effect is canceled in the $S$-matrix by appropriate corrections to the LSZ formula~\cite{Lehmann:1954rq}. The transformation~\eqref{Vg_transf} defines then an unobservable redundancy of description in these processes. 

At the operator level, this implies that near-Z-pole data cannot probe effective operators that correct the kinetic terms of the $W$ and $B$ fields. In the Warsaw basis these are $Q_{HW}$ and $Q_{HB}$, which in fact do not contribute to the scattering amplitudes. 
In addition, because $S$-matrix elements are invariant under field transformations consistent with Equations of Motion (EOM) relations, \twotwo processes are also blind to linear combinations of other operators that are equivalent to $Q_{HW},\,Q_{HB}$ via EOM. These relations can be written \cite{Kilian:2003xt,Grojean:2006nn,Grzadkowski:2010es,Alonso:2013hga}
 \begin{align}
 \hyp_h\, Q_{HB}&= 
 -\frac{Q_{Hd}}{3}-Q_{He}-\frac{Q_{Hl}^{(1)}}{2}+\frac{Q_{Hq}^{(1)}}{6}+\frac{2}{3}Q_{Hu}
 +2Q_{HD}-\frac{Q_{HWB}}{2t_{\hat\theta}} +\frac{Q_{H\square}}{2}
 +\frac{2i}{g_1} (D^\mu H)^\dag (D^\nu H) B_{\mu\nu},\nonumber\\
\frac{1}{2}Q_{HW} &= 
\frac{Q_{Hq}^{(3)}+Q_{Hl}^{(3)}-t_{\hat\theta}Q_{HWB}+Q_{H\square}}{2}
+2H^\dag H(D_\mu H^\dag D^\mu H)
+\frac{2i}{g_2} (D^\mu H)^\dag \tau^i (D^\nu H)  W_{\mu\nu}^I \label{EOMrelations_complete},
 \end{align}
where $g_{1,2}$ are the $U(1)_Y$ and $SU(2)_L$ coupling constants and $\tau^I$ are the Pauli matrices. The tangent of the Weinberg angle is denoted by $t_{\hat\theta}$, and
$\hyp_h=1/2$ is the hypercharge of the $H$ field. 
Projecting Eqs.~\eqref{EOMrelations_complete} onto \twotwo matrix elements and translating to the Wilson coefficients space, one finds that the following vectors cannot be constrained by this class of processes:
\begin{equation}\label{wBW}
\begin{aligned}
 w_B &=
 -\frac{1}{3}C_{Hd}- C_{He}-\frac{1}{2} \CHls+\frac{1}{6} \CHqs+\frac{2}{3} C_{Hu}+2C_{HD}-\frac{1}{2t_{\hat\theta}}C_{HWB},\\
 w_W &= \frac{\CHqt+ \CHlt}{2}-\frac{t_{\hat\theta}}{2}C_{HWB}.
\end{aligned}
\end{equation}
The combinations $w_B$ and $w_W$ constitute a basis for the vector space of unconstrained directions: in fact $w^\a_{1,2}$ can be decomposed as ${w^{\a}_1= -w_B - 2.59 w_W}$, $w^{\a}_2= -w_B +4.31  w_W$. Analogous expressions are found for $w_{1,2}^{m_W}$~\cite{Brivio:2017bnu}, indicating that the origin of the unconstrained directions is not related to the choice of input parameters.

The reason why the inclusion of \twofour processes breaks the degeneracy is related to these scatterings allowing diagrams with triple gauge vertices (TGC), such as that in Figure~\ref{fig:diagrams}, right. In the SMEFT, these interactions generally receive contributions often labeled $g_1^{Z,\gamma},\,\kappa_{Z,\gamma},\,\lambda_{Z,\gamma}$, that cannot be interpreted as rescalings of the $W$ or $B$ fields. The presence of these terms explicitly breaks the reparameterization invariance. At the operator level, this can be understood noting that \twofour processes are sensitive to the terms $(D_\mu H)^\dagger \tau^I (D_\nu H) W_I^{\mu\nu}$, $(D_\mu H)^\dagger (D_\nu H) B^{\mu\nu}$ appearing in Eqs.~\eqref{EOMrelations_complete}, that give TGC corrections. Because the Warsaw basis does not contain these invariants, their contributions are rather expressed by the combinations of Wilson coefficients $w_W,\,w_B$, that can now be constrained. Thus the flat directions are lifted.

The reparameterization invariance discussed here is a property of \twotwo processes and not of the SMEFT. This implies, in particular, that the results presented are independent of the operator basis adopted. Nonetheless, the presence of the two unconstrained directions can emerge differently using bases other than the Warsaw one. For instance, if two of the fermionic invariants appearing in Eqs.~\eqref{EOMrelations_complete} are traded for  $(D_\mu H)^\dagger \tau^I (D_\nu H) W_I^{\mu\nu}$ and $(D_\mu H)^\dagger (D_\nu H) B^{\mu\nu}$, then the two unconstrained quantities in \twotwo data are just the Wilson coefficients associated to the latter operators (together with $C_{HW},\, C_{HB}$), rather than the vectors $w_{1,2}^\a$. Other approaches to the global analysis may also hide the presence of unconstrained directions: this happens, for instance, if the Wilson coefficients are replaced with a non  gauge-invariant parameterization in the fit, because the EOM information is typically lost in these cases.

\providecommand{\href}[2]{#2}\begingroup\raggedright\endgroup


\begin{thebibliography}{10}

\bibitem{Brivio:2017btx}
I.~Brivio, Y.~Jiang and M.~Trott, \emph{{The SMEFTsim package, theory and
  tools}},  \href{https://arxiv.org/abs/1709.06492}{{\tt 1709.06492}}.

\bibitem{Grzadkowski:2010es}
B.~Grzadkowski, M.~Iskrzynski, M.~Misiak and J.~Rosiek, \emph{{Dimension-Six
  Terms in the Standard Model Lagrangian}},
  \href{http://dx.doi.org/10.1007/JHEP10(2010)085}{\emph{JHEP} {\bf 1010}
  (2010) 085}, [\href{https://arxiv.org/abs/1008.4884}{{\tt 1008.4884}}].

\bibitem{Z-pole}
T.~ALEPH, DELPHI, L3, OPAL, S.~Collaborations, the LEP Electroweak
  Working~Group et~al., \emph{{Precision Electroweak Measurements on the Z
  Resonance}}, {\emph{Phys. Rept.} {\bf 427} (2006) 257},
  [\href{https://arxiv.org/abs/hep-ex/0509008}{{\tt hep-ex/0509008}}].

\bibitem{Falkowski:2014tna}
A.~Falkowski and F.~Riva, \emph{{Model-independent precision constraints on
  dimension-6 operators}},
  \href{http://dx.doi.org/10.1007/JHEP02(2015)039}{\emph{JHEP} {\bf 02} (2015)
  039}, [\href{https://arxiv.org/abs/1411.0669}{{\tt 1411.0669}}].

\bibitem{Berthier:2015oma}
L.~Berthier and M.~Trott, \emph{{Towards consistent Electroweak Precision Data
  constraints in the SMEFT}},
  \href{http://dx.doi.org/10.1007/JHEP05(2015)024}{\emph{JHEP} {\bf 05} (2015)
  024}, [\href{https://arxiv.org/abs/1502.02570}{{\tt 1502.02570}}].

\bibitem{Berthier:2015gja}
L.~Berthier and M.~Trott, \emph{{Consistent constraints on the Standard Model
  Effective Field Theory}},
  \href{http://dx.doi.org/10.1007/JHEP02(2016)069}{\emph{JHEP} {\bf 02} (2016)
  069}, [\href{https://arxiv.org/abs/1508.05060}{{\tt 1508.05060}}].

\bibitem{Grinstein:1991cd}
B.~Grinstein and M.~B. Wise, \emph{{Operator analysis for precision electroweak
  physics}},
  \href{http://dx.doi.org/10.1016/0370-2693(91)90061-T}{\emph{Phys.Lett.} {\bf
  B265} (1991) 326--334}.

\bibitem{Han:2004az}
Z.~Han and W.~Skiba, \emph{{Effective theory analysis of precision electroweak
  data}}, \href{http://dx.doi.org/10.1103/PhysRevD.71.075009}{\emph{Phys.Rev.}
  {\bf D71} (2005) 075009}, [\href{https://arxiv.org/abs/hep-ph/0412166}{{\tt
  hep-ph/0412166}}].

\bibitem{Grojean:2006nn}
C.~Grojean, W.~Skiba and J.~Terning, \emph{{Disguising the oblique
  parameters}}, \href{http://dx.doi.org/10.1103/PhysRevD.73.075008}{\emph{Phys.
  Rev.} {\bf D73} (2006) 075008},
  [\href{https://arxiv.org/abs/hep-ph/0602154}{{\tt hep-ph/0602154}}].

\bibitem{Brivio:2017bnu}
I.~Brivio and M.~Trott, \emph{{Scheming in the SMEFT... and a
  reparameterization invariance!}},
  \href{http://dx.doi.org/10.1007/JHEP07(2017)148}{\emph{JHEP} {\bf 07} (2017)
  148}, [\href{https://arxiv.org/abs/1701.06424}{{\tt 1701.06424}}].

\bibitem{Passarino:2012cb}
G.~Passarino, \emph{{NLO Inspired Effective Lagrangians for Higgs Physics}},
  \href{http://dx.doi.org/10.1016/j.nuclphysb.2012.11.018}{\emph{Nucl. Phys.}
  {\bf B868} (2013) 416--458}, [\href{https://arxiv.org/abs/1209.5538}{{\tt
  1209.5538}}].

\bibitem{deFlorian:2016spz}
{\scshape LHC Higgs Cross Section Working Group} collaboration, D.~de~Florian
  et~al., \emph{{Handbook of LHC Higgs Cross Sections: 4. Deciphering the
  Nature of the Higgs Sector}},  \href{https://arxiv.org/abs/1610.07922}{{\tt
  1610.07922}}.

\bibitem{Hagiwara:1993ck}
K.~Hagiwara, S.~Ishihara, R.~Szalapski and D.~Zeppenfeld, \emph{{Low-energy
  effects of new interactions in the electroweak boson sector}},
  \href{http://dx.doi.org/10.1103/PhysRevD.48.2182}{\emph{Phys. Rev.} {\bf D48}
  (1993) 2182--2203}.

\bibitem{DeRujula:1991ufe}
A.~De~Rujula, M.~B. Gavela, P.~Hernandez and E.~Masso, \emph{The selfcouplings
  of vector bosons: Does lep-1 obviate lep-2?},
  \href{http://dx.doi.org/10.1016/0550-3213(92)90460-S}{\emph{Nucl. Phys.}
  (1992) 3--58}.

\bibitem{Berthier:2016tkq}
L.~Berthier, M.~Bj\o rn and M.~Trott, \emph{{Incorporating doubly resonant
  $W^\pm$ data in a global fit of SMEFT parameters to lift flat directions}},
  \href{http://dx.doi.org/10.1007/JHEP09(2016)157}{\emph{JHEP} {\bf 09} (2016)
  157}, [\href{https://arxiv.org/abs/1606.06693}{{\tt 1606.06693}}].

\bibitem{Lehmann:1954rq}
H.~Lehmann, K.~Symanzik and W.~Zimmermann, \emph{{On the formulation of
  quantized field theories}},
  \href{http://dx.doi.org/10.1007/BF02731765}{\emph{Nuovo Cim.} {\bf 1} (1955)
  205--225}.

\bibitem{Kilian:2003xt}
W.~Kilian and J.~Reuter, \emph{{The Low-energy structure of little Higgs
  models}}, \href{http://dx.doi.org/10.1103/PhysRevD.70.015004}{\emph{Phys.
  Rev.} {\bf D70} (2004) 015004},
  [\href{https://arxiv.org/abs/hep-ph/0311095}{{\tt hep-ph/0311095}}].

\bibitem{Alonso:2013hga}
R.~Alonso, E.~E. Jenkins, A.~V. Manohar and M.~Trott, \emph{{Renormalization
  Group Evolution of the Standard Model Dimension Six Operators III: Gauge
  Coupling Dependence and Phenomenology}},
  \href{http://dx.doi.org/10.1007/JHEP04(2014)159}{\emph{JHEP} {\bf 1404}
  (2014) 159}, [\href{https://arxiv.org/abs/1312.2014}{{\tt 1312.2014}}].

\end{thebibliography}
\end{document}